# Reference-free polarization-sensitive quantitative phase imaging using singe-point optical phase conjugation

SEUNGWOO SHIN,[1,2] KYEOREH LEE,[1,2] ZAHID YAQOOB,[3] PETER T.C. SO,[3,5] AND YONGKEUN PARK[1,2,4,*]

[1]*Department of Physics, Korea Advanced Institute of Science and Technology (KAIST), Daejeon 34141, Republic of Korea*
[2]*KAIST Institute for Health Science and Technology, KAIST, Daejeon 34141, Republic of Korea*
[3]*Laser Biomedical Research Center, G. R. Harrison Spectroscopy Laboratory, Massachusetts Institute of Technology, Cambridge, Massachusetts, 02139, USA*
[4]*TomoCube Inc., Daejeon 34051, Republic of Korea*
[5]*E-mail: ptso@mit.edu*
*\*Corresponding author: yk.park@kaist.ac.kr*

**We propose and experimentally demonstrate a method of polarization-sensitive quantitative phase imaging using two photo detectors. Instead of recording wide-field interference patterns, finding the modulation patterns maximizing focused intensities in terms of the polarization states enables polarization-dependent quantitative phase imaging without the need for a reference beam and an image sensor. The feasibility of the present method is experimentally validated by reconstructing Jones matrices of various samples including a polystyrene microsphere, a maize starch granule, and a rat retinal nerve fiber layer. Since the present method is simple and sufficiently general, we expect that it may offer solutions for quantitative phase imaging of birefringent materials.**

Light waves can be completely described by three quantities of an electric field vector; a temporal frequency $\omega$, complex amplitude or an optical field $E$, and a polarization state $p$. Depending on which of the three quantities are to be measured, various optical measurement methodologies may be divided into several types. Polarization microscopy refers to measuring absolute values of the polarization states of the optical field from light scattered off a sample [1, 2]. Holography allows the measurement of an optical field $E(x, y)$ by recording interference patterns with a well-defined reference beam [3-5].

By combining holography and polarization microscopy, optical fields depending on polarization states $E(x, y)_p$ can be measured [6-8]. Polarization holographic microscopic or polarization-sensitive quantitative phase imaging (QPI) techniques have been successfully demonstrated in measuring scattered fields from various birefringent samples [7-11]. However, applications of the techniques have been restricted due to the limitations of complicated optical setups, which require precise alignment and specific wavelengths due to the use of 2D image sensors.

Unfortunately, QPI or holographic imaging at visible wavelengths provides limited molecular information for cellular biomolecules such as proteins, carbohydrates, lipids, and small molecules [12-14]. Although refractive index (RI) distributions inside a cell can be obtained using various QPI imaging methods, different species of biomolecules have limited distinguishability via such properties [14, 15]. Since RIs of biomolecules show high molecular specificity at ultraviolet (UV) or infrared (IR) wavelengths, molecular-specific measurements have been conducted at these wavelengths, for example, using coherent anti-Stokes Raman spectroscopy [16], and UV and vibrational circular dichroism (CD) spectroscopy [17, 18]. Therefore, to realize molecular-specific polarization holographic imaging, it is crucial to measure UV or IR optical fields, overcoming the limitations of available image sensors.

Recently, to remedy the limited availability of image-sensors in the invisible wavelength regime, single-pixel cameras have been employed for holographic imaging [19, 20]. From intensity-correlation measurements, single-pixel cameras provide only intensity images using a single photodiode [21, 22]. By replacing an image sensor with a single-pixel camera in an interferometer, an optical field can be obtained using a single photodiode. However, such an approach shows inferior imaging performance due to the inevitable phase fluctuations and time-consuming intensity-correlation measurements [23].

Instead of employing a single-pixel camera in an interferometric setting, an alternative approach called single-point holography [23], was presented by exploiting the time-reversal nature of optical phase conjugation [24, 25]. When a scattered field $S$ is modulated by its complex conjugate $S^*$, the scattered field becomes a plane wave which generates a maximum intensity at the focal plane of a focusing lens (Fig. 1). Contrary to conventional holographic imaging, single-point holography allows the measurement of an optical field by finding the pattern that maximizes the focused intensity after a lens, instead of recording an interference pattern on an image sensor or a single-pixel camera.

Here, by expanding the concept of single-point holography, we propose and experimentally demonstrate a method for polarization-sensitive holographic microscopy without the use of an image sensor (Fig. 1). By employing polarization-dependent point measurements, the patterns maximizing the focused intensities corresponding to the polarization states can be found. From the obtained patterns, optical fields corresponding to the polarization states can be determined.

The experimental setup is shown in Fig. 2(a). To achieve fast and polarization-independent modulations, a digital micromirror device (DMD, V-7001, Vialux Inc.) is utilized. A linearly polarized plane wave ($\lambda$

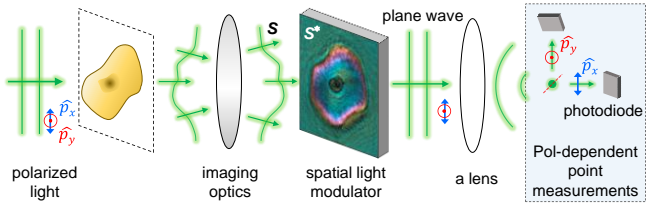

**Fig. 1.** Schematic illustration of the present method. Under a polarized plane-wave illumination, the scattered light $S$ from a sample is relayed onto a spatial light modulator (SLM) by imaging optics. When the SLM displays conjugated phase of the scattered light $S^*$, the focused light shows the maximum intensity at a point. Polarization- (Pol)-dependent point measurements find the optical phase conjugation patterns for both orthogonal polarization states.

= 532 nm) impinges onto a sample via a polarizer, a lens (L1, $f$ = 200 mm), and a condenser lens (W Plan-Apochromat, 63×, NA = 1.0, Carl Zeiss AG). The scattered field from the sample is collected and relayed onto the DMD by an objective lens (MPLAPON-Oil, 100×, NA = 1.4, Olympus Inc.), a tube lens within a conventional microscope ($f$ = 200 mm, TE 2000-E, Nikon Inc.), and additional 4-$f$ relay lenses (L2–3, $f$ = 200 mm) with spatial filtering for corresponding NA of 0.93. This spatial filtering enables the phase modulation using a DMD, which is minutely described in the following paragraph. Then, the relayed scattered field is modulated by the DMD, and a polarization-maintaining single mode fiber transmits the focused light after a lens (L4, $f$ = 25 mm) to polarization-sensitive point detection, which consists of a polarizing beamsplitter and two avalanche photodiodes (APDs).

To effectively find the pattern maximizing focused intensity after a lens, binary patterns displayed on the DMD are constructed using phase shifts and a Hadamard basis, which have also been used elsewhere [Fig. 2(b)] [23]. For phase modulation using a DMD, neighboring 2×2 DMD pixels having different phase allocations are optically combined by the NA of imaging optics [26]. Then, by analyzing recorded intensities corresponding to the displayed patterns, we can find the optical phase conjugation pattern [23]. Since a DMD modulates light regardless of its incident polarization state, the optical phase conjugation patterns with respect to the polarization states can be simultaneously found from the measured intensities using two APDs [Fig. 2(c)]. In this work, optical resolution of the measured field is 350 nm, and the measurement time is 4.9 seconds for a DMD display rate of 10 kHz.

For experimental validation of the proposed method, a spatially resolved Jones matrix of a sample is reconstructed. To reconstruct a spatially resolved Jones matrix of a sample, scattered fields from a sample are measured with polarization sensitivity using the proposed method, corresponding to two different polarization states of illuminations, $p_+ = (p_x + p_y)/\sqrt{2}$ and $p_- = (p_x - p_y)/\sqrt{2}$, respectively. Then, the polarization-dependent measured fields using the APDs can be expressed as $(\mathbf{J}_{11} \pm \mathbf{J}_{12})/\sqrt{2}$ and $(\mathbf{J}_{21} \pm \mathbf{J}_{22})/\sqrt{2}$ with respect to the illumination $p_\pm$, where $\mathbf{J}_{mn}$ indicates a matrix component of a Jones matrix ($m, n \in \{1, 2\}$). Thereby, the Jones matrix is reconstructed via linear arithmetic calculations on the measured fields.

To experimentally demonstrate the proposed method, spatially resolved Jones matrices of various samples are reconstructed [8, 10, 11, 27, 28]; a non-birefringent sample [a 10–μm–diameter polystyrene (PS) microsphere immersed in oil ($n$ = 1.56), Fig. 3(a)] and a birefringent sample [a maize starch granule (Corn Starch, Argo®) immersed in oil ($n$ = 1.48), Fig. 3(b)], respectively. As expected, contrary to the diagonal

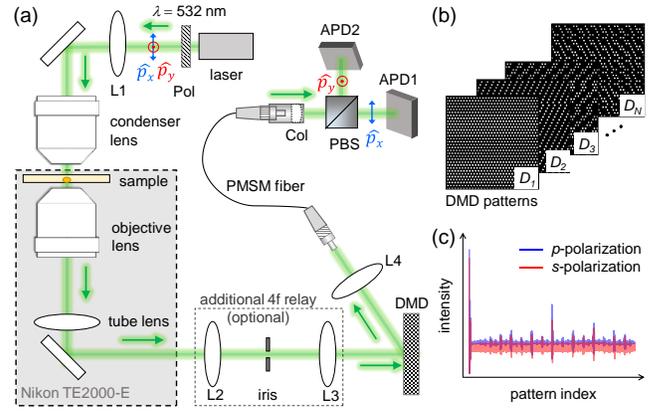

**Fig. 2.** (a) Experimental setup. $p_x$ and $p_y$ denote linear polarization states whose polarization directions are parallel to the $x$- and $y$-axes, respectively. Pol, polarizer; L1–L4, lenses ($f_{1-3}$ = 200 mm; $f_4$ = 25 mm); [DMD, digital micromirror device; PMSM fiber, polarization-maintaining single mode fiber; Col, collimation lens ($f$ = 15 mm); PBS, polarizing beamsplitter; APD, avalanche photodiode]. (b) Sequentially displayed binary patterns on the DMD and (c) polarization-dependent measured intensities using the APDs.

Jones matrix of the non-birefringent PS microsphere, the Jones matrix of the maize starch granule has off-diagonal terms which demonstrate the birefringence of the starch granule [Fig. 3].

To demonstrate the applicability of the present method, we systematically analyze anisotropic optical properties of a maize starch granule using its reconstructed Jones matrix [Fig. 4]. Maize starch granules have strong birefringence and optical activity due to the two major molecular constituents, namely, amylose (28.7%) and amylopectin (71.3%) [29]. In particular, amylopectin, that has a double helix molecular structure, forms extensively branched supramolecular crystallite structure in a starch granule [Fig. 4(a)] [30-32]. From a hilum inside the starch granule, branched chains of amylopectin grow in radial directions [Fig. 4(a)] [32]. Due to the chiral molecular and the supramolecular structure, a maize starch granule shows distinctive anisotropic optical properties which have been investigated by various methods including polarized light microscopy [31], second harmonic generation microscopy [33-35], and CD spectroscopy [35, 36].

The Jones matrix of a maize starch granule can provide scattered fields from the granule for any incident polarization state, which allows us to investigate anisotropic optical properties of the starch including retardance, elliptical polarization states of eigenstates, and linear and circular dichroism [10, 28, 37] [Figs. 4(b)–4(f)]. To systematically analyze the anisotropic optical properties, the spatially resolved Jones matrix is diagonalized as,

$$\mathbf{J} q_i = \mu_i q_i, \quad i \in \{1, 2\}, \tag{1}$$

where $q_i$ are the eigenstates of the Jones matrix $\mathbf{J}$, and $\mu_i$ are the corresponding eigenvalues. Since an incident polarization state as an eigenstate is not changed but the amplitude and phase of the light is changed as $\mu_i$, after scattering by the sample, the polarization eigenstates correspond to the fast or slow axes [10, 28, 37].

From the spatially resolved eigenvalues, we can obtain the retardance, a measure of the difference of optical paths with respect to the polarization eigenstates, as $R = (\angle \mu_2 - \angle \mu_1) \times \lambda / 2\pi$ [Fig. 4(b)].

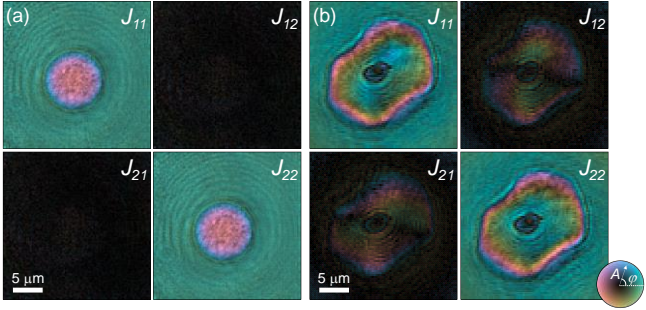

**Fig. 3.** Reconstructed Jones matrices of (a) a 10–μm–diameter polystyrene microsphere immersed in oil and (b) a maize starch granule immersed in oil. The superscript $mn$ in $\mathbf{J}_{mn}$ indicates the corresponding matrix component of the spatially resolved Jones matrix ($m, n \in \{1, 2\}$). In the color circle, overlaid at the lower right, the symbols $A$ and $\varphi$ denote the normalized amplitude and phase, respectively.

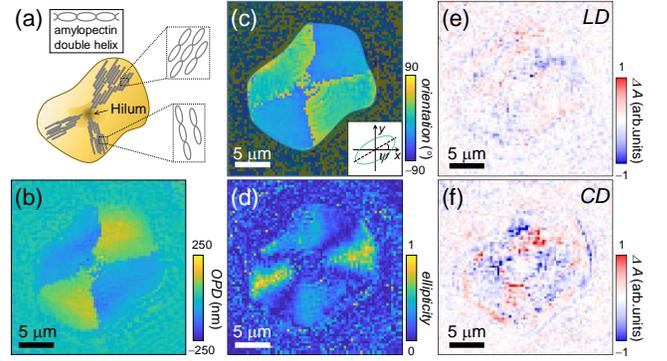

**Fig. 4.** (a) Schematic illustration of a helical molecular structure and a supramolecular crystallite structure of amylopectin inside a maize starch granule. Various anisotropic optical properties of the starch granule can be investigated using its reconstructed Jones matrix. (b) Spatially resolved retardance and (c) polarization orientations of a polarization eigenstate with respect to the $x$-axis. The color scale represents the various polarization orientations. (d) The ellipticity of polarization eigenstates, and (e) linear and (f) circular dichroism. [OPD, optical path difference; LD, linear dichroism; CD, circular dichroism].

Orientation and shape of a polarization ellipse of an eigenstate $q_1$ can be visualized using angles between the major axis of the ellipse and the $x$-axis [Fig. 4(c)], and the ellipticity which is the ratio of the minor to major axes of the ellipse [Fig. 4(d)]. In the Fig. 4(c), due to the coherent noise of a laser, eigenstates in the absence of a sample show linear polarization states with various orientations, which are shaded to highlight information in the presence of a sample.

Also, using scattered fields under different linear polarized illumination, we can obtain linear dichroism (LD), which is a measure of absorption difference for different orientations of incident linear polarizations [Fig. 4(e)]. Similarly, CD, a measure of absorption difference for incident left-handed and right-handed circular polarizations, can also be obtained using the Jones matrix [Fig. 4(f)], $\Delta A = A_L - A_R$, where $A_L$ and $A_R$ denote absorption for incident left-handed, $p_L = (p_x + ip_y)/\sqrt{2}$, and right-handed, $p_R = (p_x - ip_y)/\sqrt{2}$, circular polarizations, respectively.

The radially grown supramolecular crystallite structures of amylopectin inside a maize starch granule [Fig. 4(a)] have azimuthally varying anisotropic optical properties, and the various analyses using the Jones matrix [Figs. 4(b)–4(f)] give consistent results. In particular, the helical molecular structure and the supramolecular structures of amylopectin induce significant CD signals which can be obtained using the proposed method. For the molecular-specific investigation, CD measurements have been conducted using invisible wavelengths such as UV and IR [17, 18]. Because the present method does not require an image sensor and an interferometer for holographic imaging [23], the present method can be employed for invisible holographic imaging for molecular-specific investigation of chiral molecules.

To further demonstrate the applicability of the present method, we also measured a Jones matrix of a fixed rat retinal nerve fiber layer [Figs. 5(a) and 5(b)] [38, 39]. Similar to the measurement of a maize starch granule, a Jones matrix of a part of the retinal nerve fiber is reconstructed using the proposed method, from which the retardance and polarization ellipticity of an eigenstate can be obtained [Figs. 5(c)–5(e)]. As qualitatively indicated in cross-polarized light microscopy [Fig. 5(b)], part of the retinal nerve fiber shows distinct birefringence which can be quantitatively measured and analyzed by reconstructing its Jones matrix [38, 39].

In conclusion, we have proposed and experimentally demonstrated a method of polarization holographic microscopy without the use of an image sensor. Instead of recording interference patterns using an interferometer, finding the modulation patterns that maximize focused intensities with respect to the polarization states enables polarization-dependent holographic imaging. Feasibility of the present method is validated by reconstructing Jones matrices of various samples, including a non-birefringent microsphere, a birefringent starch granule, and a biological tissue. Using the Jones matrices, we also present systematic analyses of anisotropic optical properties of the birefringent samples. Although the present method is slow (0.5 Hz for 128 × 128 spatial modes and the maximum DMD display rate of 22.7 kHz), this disadvantage can be alleviated by employing compressive sampling algorithms or deep learning algorithms for rapid determination of the optimal pattern for maximizing focused intensity after a lens.

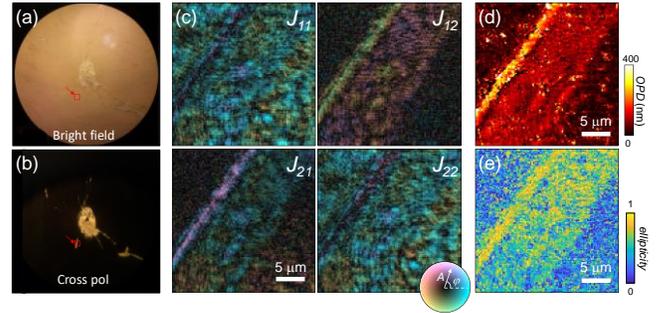

**Fig. 5.** Applications to a birefringent tissue sample. (a) Bright-field and (b) cross-polarized light microscopic images which show the entire sliced rat retinal nerve fiber tissue and its birefringence. (c) Spatially resolved Jones matrix, (d) retardance, and (e) polarization ellipticity of an eigenstate of the region of the retinal nerve fiber layer indicated by red arrows in (a) and (b).

It is noteworthy that the present method can be readily expanded to polarization-dependent holographic imaging in the invisible wavelengths, which may allow molecular-specific optical measurement of chiral molecules. Furthermore, by replacing the photodiodes in the present method with spectrometers or the monochromatic illumination source with a wavelength scanning source, it would also enable full optical measurement corresponding to the three light properties: temporal frequency, complex amplitude, and polarization states. This expansion of the present method using spectrometers can provide invaluable information about a sample, including anisotropic optical properties with dispersion [40].

**Acknowledgement.** The author wishes to thank Baoliang Ge and Renjie Zhou, for providing a tissue sample.

**Funding.** This work was supported by KAIST, BK21+ program, Tomocube, National Institutes of Health (NIH) (1R01EY017656-06, 1U01NS090438-01, 9P41EB015871-28, 1-R01HL121386-01), and National Research Foundation of Korea (2017M3C1A3013923, 2015R1A3A2066550, 2014K1A3A1A09063027).